\documentclass[a4paper,10pt]{article}
\usepackage{graphicx,amsmath,natbib}
\def\h0units{\mathrm{km\,s^{-1}\,Mpc^{-1}}}
\def\apj{ApJ\,  }

\def\pla{Phys. Lett. A   }
\def\physa{Phys. A    }

\def\za{Z. Astrophys.  } 
\title
{
The transition from order to disorder in  Voronoi Diagrams
with applications
}
\author{L. Zaninetti   \\ 
Department of Physics,
 via P.Giuria 1,\\ I-10125 Turin,Italy   \\
\footnote{zaninetti@ph.unito.it}\hspace{0.08cm}
{Corresponding author: zaninetti@ph.unito.it}}
\begin{document}
\maketitle
\begin{abstract}
The transition from ordered to disordered structures in
Voronoi tessellation
is obtained by perturbing  the seeds
that were originally identified with two types of lattice
in 2D   and one type in 3D.
The  area  in 2D and the volume in 3D are modeled with
the Kiang function.
A new relationship  that models  the scaling  of the Kiang
function with   a geometrical parameter is introduced.
A first  application
models  the local structure of sub- and supercritical ammonia
as function of the temperature
and a  second application models
the volumes of cosmic voids.
\end {abstract}

\section{Introduction}

The seeds in 1D, 2D and 3D Voronoi Diagrams
are usually taken to be randomly distributed,
the so called
Poisson-Voronoi tessellations
(PVT)
\cite{Okabe2000}.
A first generalization of the PVT
are the quasi random seeds,
such as   the Sobol seeds
\cite{Sobol1967,Bratley1988,Zaninetti1992,Zaninetti2009c,Zaninetti2015b}
and the eigenvalues
of complex random matrices \cite{Lecaer1990}.
A second  major generalisation
modifies regular structures
to produce non-Poissonian seeds
for the Voronoi tessellation (NPVT).
We select some methods, including
perturbation of cubic structures \cite{Lucarini2009},
generation of seeds  with controlled regularity \cite{Zhu2014},
an information geometric model to simulate  graphene \cite{Dodson2015},
a 3D topological analysis  \cite{Lazar2015}
and
two-dimensional perturbed systems \cite{Leipold2016}.

The rest of this paper is organised as follows:
In section \ref{useful}, we first review the existing formulae
which  model the areas and volumes in 2D/3D
PVT.
Afterwards,  we introduce two models for NPVT, see Section
\ref{nonpoisson}.
Two applications are discussed in Section \ref{applications}.

\section{Useful formulae in PVT}
\label{useful}

\begin{figure}
\begin{center}
\includegraphics[width=10cm]{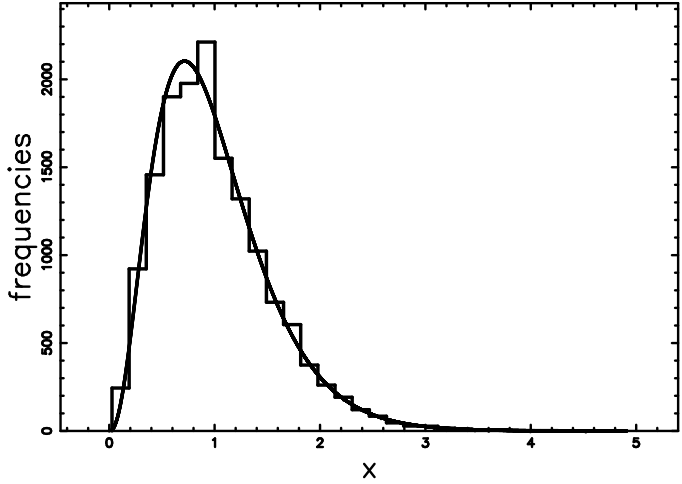}
\end {center}
\caption
{
Histogram (step-diagram) of the 2D PVT
reduced  area  distribution.
}
 \label{gamma_kiang_2d_poisson}%
 \end{figure}

The  probability density function (PDF) for  
segments (1D) in  PVT
is modeled by
a gamma variate
\begin{equation}
 H (x ;c ) = \frac {c} {\Gamma (c)} (cx )^{c-1} \exp(-cx)
\quad ,
\label{kiang}
\end{equation}
where $ 0 \leq x < \infty $ , $ c~>0$
and $\Gamma (c)$ is the gamma function with argument c,
see  formula~(5) in \cite{kiang}.
In the case of 1D, c=2  which is an analytical result.
Conversely  the  PDF for areas in 2D and volumes
in 3D
was  conjectured  to follow the above gamma variate
with c=4 and  c=6.
Later on, the   "Kiang conjecture"
was  refined by  \cite{Ferenc_2007}
with the following PDF
\begin{equation}
FN(x;d) = Const \times x^{\frac {3d-1}{2} } \exp{(-(3d+1)x/2)}
\quad ,
\label{rumeni}
\end{equation}
where
\begin{equation}
Const =
\frac
{
\sqrt {2}\sqrt {3\,d+1}
}
{
2\,{2}^{3/2\,d} \left( 3\,d+1 \right) ^{-3/2\,d}\Gamma \left( 3/2\,d+
1/2 \right)
}
\quad ,
\end{equation}
and $d(d=1,2,3)$ represents the
dimension of the considered space.
This  PDF
allows us to fix the
"Kiang conjecture" in
c=3.5 and c=5   for the 2D and 3D PVT case.
A typical 2D result   is reported in 
Figure \ref{gamma_kiang_2d_poisson}
for the reduced area distribution
when the Poissonian seeds are 20000, $d=2$
and $c=3.50$.

\begin{figure}
\begin{center}
\includegraphics[width=10cm]{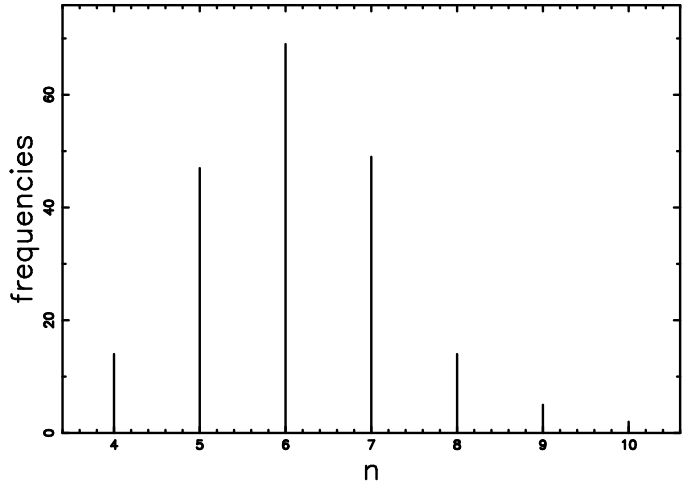}
\end {center}
\caption
{
Histogram
of the expected  number of edges in the 2D PVT.
}
 \label{lati2dpoisson}%
 \end{figure}

The theoretical expected  number of edges of a typical 2D cell
is six, see Table 5.5.1 in \cite{Okabe2000} and our
results
are reported in Figure \ref{lati2dpoisson},
in which the averaged edges are 6.125.

\begin{figure}
\begin{center}
\includegraphics[width=10cm]{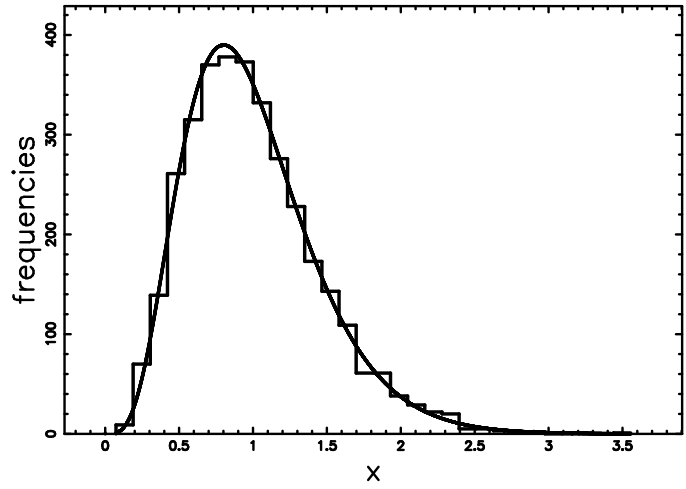}
\end {center}
\caption
{
Histogram (step-diagram) of the 3D PVT
reduced  volume  distribution.
}
 \label{gamma_kiang_3d}%
 \end{figure}

\begin{figure}
\begin{center}
\includegraphics[width=10cm]{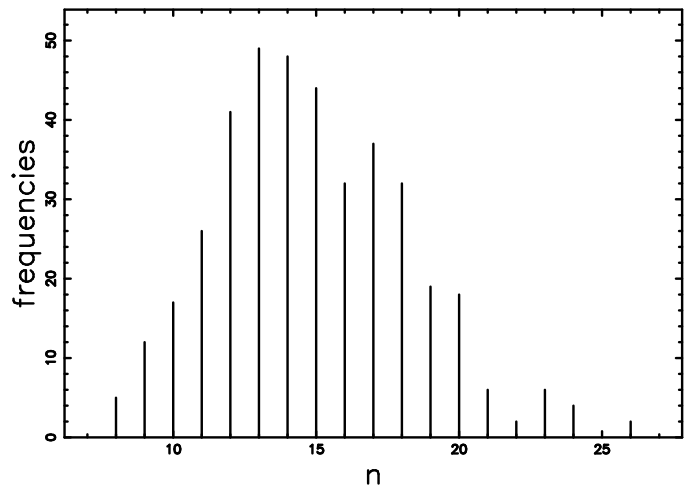}
\end {center}
\caption
{
Histogram
of the expected  number of  faces in the 3D PVT.
}
 \label{numberfaces}%
 \end{figure}

In the 3D case  Figure \ref{gamma_kiang_3d} reports
the histogram for the volume distribution
when the Poissonian seeds are 4096, $d=3.03$
and $c=5.05$.
and Figure \ref{numberfaces} the histogram of the number of faces
with average value 14.897. A comparison should be done
with the expected value   15.535, see
Table 5.5.2 in \cite{Okabe2000}.

\subsection{The lognormal distribution}

The lognormal PDF, $f_{LN}$,
is
\begin{equation}
f_{LN} (x;m,\sigma) = \frac
{
{{\rm e}^{-\,{\frac {1}{{2\,\sigma}^{2}} \left( \ln  \left( {
\frac {x}{m}} \right )  \right ) ^{2}}}}
}
{
x\sigma\,\sqrt {2\,\pi}
}
\quad,
\label{pdflognormal}
\end{equation}
where $m$ is the median and $\sigma$ the shape parameter,
see \cite{evans}.
The distribution function (DF), $F_{LN}$, is
\begin{equation}
F_{LN} (x;m,\sigma) =
\frac{1}{2}+\frac{1}{2}\,{\rm erf} \left(\frac{1}{2}\,{\frac {\sqrt {2} \left( -\ln  \left( m
 \right ) +\ln  \left( x \right )  \right ) }{\sigma}}\right )
\quad ,
\label{dflognormal}
\end{equation}
where ${\rm erf(x)}$ is the error function, defined as
\begin{equation}
\mathop{\mathrm{erf}\/}\nolimits
(x)=\frac{2}{\sqrt{\pi}}\int_{0}^{x}e^{-t^{2}}dt
\quad ,
\end{equation}
see \cite{Abramowitz1965}.

\section{Non-Poissonian case}
\label{nonpoisson}

Two regular geometrical models are perturbed
to have NPVT
and a function which models the transition from
order to disorder is introduced.

\subsection{The modified lattice case}

To have more  flexible  seeds we   introduce
the  adjustable  non Poissonian seeds   (LNPVT),
which  can be computed both in 2D and 3D
following an algorithm introduced in \cite{Zaninetti2015b}.
The algorithm is now outlined:
\begin{enumerate}
\item  The process starts by  inserting the seeds
       on a 2D/3D regular
       Cartesian grid with equal distance $\delta$ between
       one point and  the following one.
\item  A random radius is generated according to the
       half Gaussian ,$HN(x)$, which  is defined
       in the interval $[0,\infty]$
\begin{equation}
HN(x;s) =
\frac {2} {s (2 \pi)^{1/2}} \exp ({- {\frac {x^2}{2s^2}}} )
\quad
0 < x < \infty
\quad .
\label{gaussianhalf}
\end{equation}
A random direction is chosen in 2D/3D and the two/three
Cartesian coordinates  of  the generated radius
are evaluated.
These two/three   small Cartesian components
are  added to the regular 2D/3D grid  which represent
the seeds.
To have  small corrections, we express
$s$  in $\delta$ units.
\item
We now  have $N$ seeds
and we can  eliminate  a given number of seeds  , $N_{hole}$,
according to the rule  $N_{hole}=N \times p_{hole}$.
This elimination of seeds will allow a more disordered
distribution of areas/volumes; otherwise specified
$p_{hole}=0$.
\end{enumerate}

\begin{figure}
\begin{center}
\includegraphics[width=10cm]{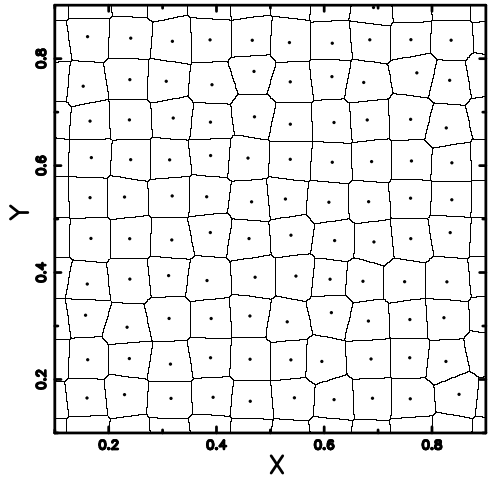}
\end {center}
\caption
{
An example of 2D LNPVT.
}
\label{area2d_lattice}
\end{figure}

Figure
\ref{area2d_lattice}
reports an example
of 2D tessellation from LNPVT in   which
$s=0.1$ and we have  400 seeds.

\begin{figure}
\begin{center}
\includegraphics[width=10cm]{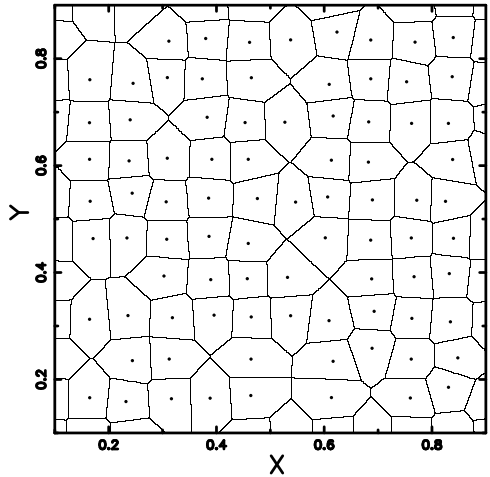}
\end {center}
\caption
{
An example of 2D LNPVT.
}
\label{area2d_lac_lattice}
\end{figure}

An example of the presence of holes in this network for  the
seeds is reported in Figure \ref{area2d_lac_lattice}
where   original seeds are 400 , $s=0.1$  and $p_{hole}=0.19$.

\begin{figure}
\begin{center}
\includegraphics[width=10cm]{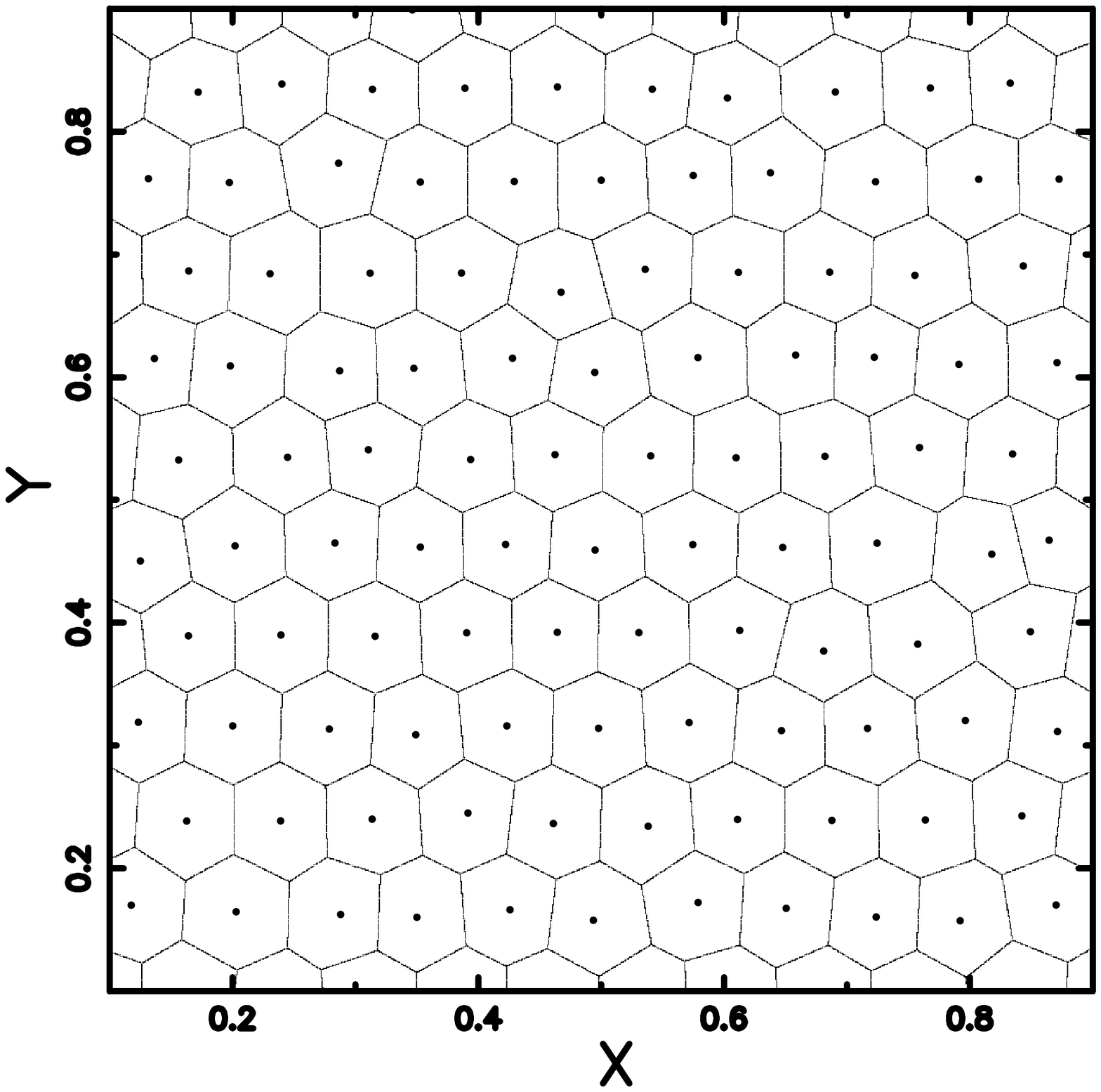}
\end {center}
\caption
{
An example of 2D TNPVT.
}
\label{area2d_triangular}
\end{figure}

Another case   uses  the  lattice points  shifted of
$delta/2$ between one row and the following row as
a first point of the suggested algorithm,
we define it as triangular non Poissonian tessellation, TNPVT.
This tessellation will be produced in 2D irregular
hexagons, see  Figure \ref{area2d_triangular}
where we have   400 seeds and $s=0.1$.

\subsection{Order-disorder transition}

\begin{figure}
\begin{center}
\includegraphics[width=10cm]{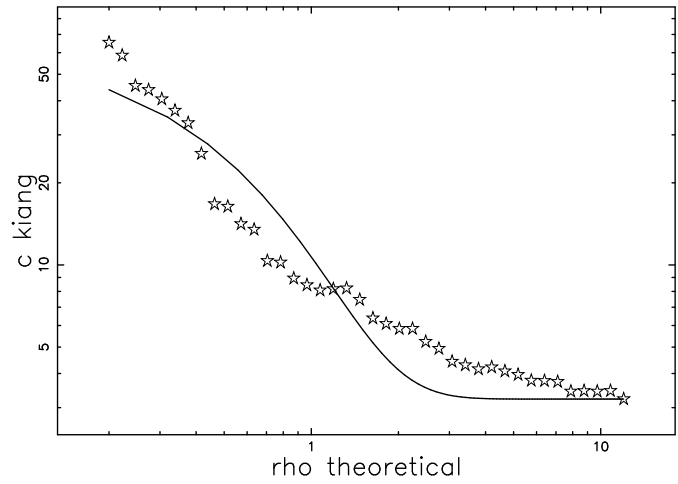}
\end {center}
\caption
{
The Kiang parameter $c$   as function of $s$
for 2D LNPVT.
}
\label{ckiang_exp}
\end{figure}

\begin{figure}
\begin{center}
\includegraphics[width=10cm]{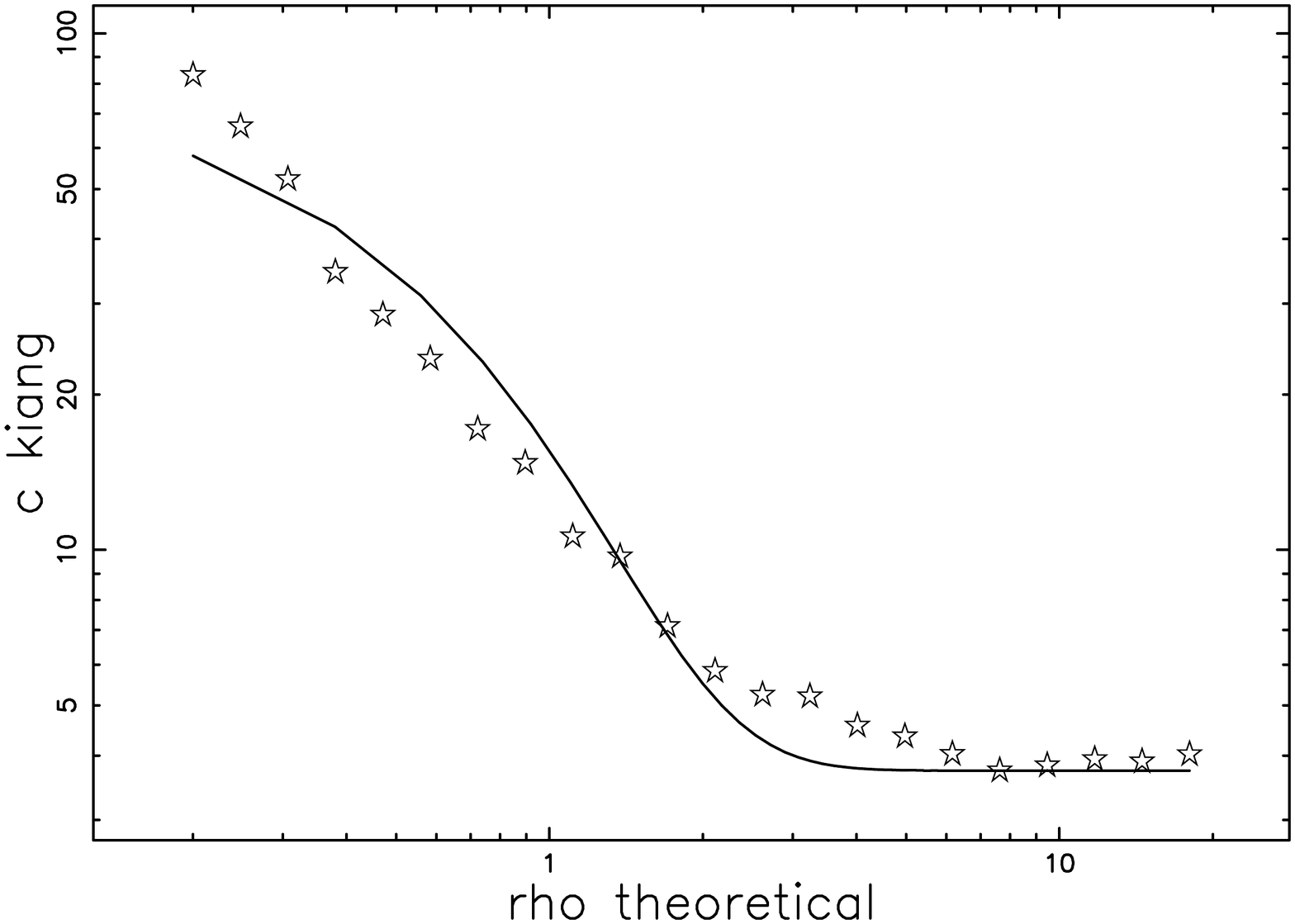}
\end {center}
\caption
{
The Kiang parameter $c$   as function of $s$
for 2D TNPVT.
}
\label{ckiang_triangular_exp}
\end{figure}

We can model the order disorder transition  introducing the following
function for  $c$ of Kiang  as in equation (\ref{kiang})
\begin{equation}
c(s;c_{min},c_{max},a) =
c_{{\max}}- \left( c_{{\max}}-c_{{\min}} \right)  \left( 1-{{\rm e}^{-
a\,s}} \right)
\label{cs}
\quad  ,
\end{equation}
where $s$ is the scale of
the half Gaussian, see equation (\ref{gaussianhalf}),
$c_{{\max}}$ and $c_{{\min}}$  are the maximum and minimum
value  for  $c$ of Kiang and  $a$ is a scale  to be found
from the fitting  procedure.
A typical 2D    example for $c$ as function of $s$
is reported in
Figure \ref{ckiang_exp} for the LNPVT case
where 
$c_{{\min}}=3.22$  (the minimum), 
$c_{{\max}}=65.33$ (the maximum)  and $a$=2.11
and   in
Figure \ref{ckiang_triangular_exp} for the LNPVT case
when $c_{{\min}}=3.73$, $c_{{\max}}=83.06 $  and $a$= 1.9.

\begin{figure}
\begin{center}
\includegraphics[width=10cm]{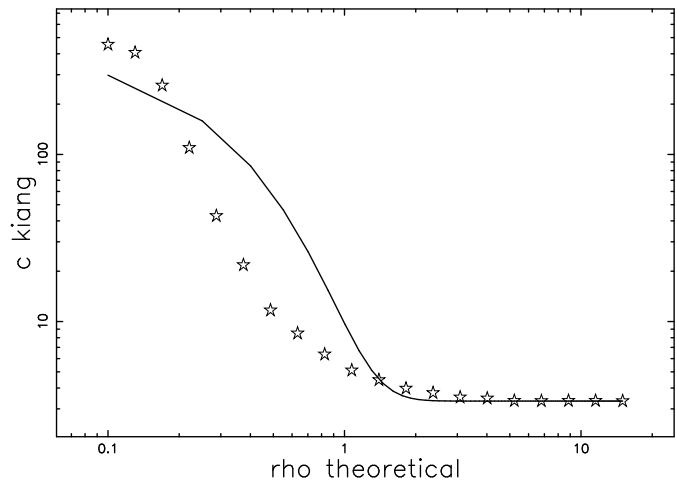}
\end {center}
\caption
{
The Kiang parameter $c$   as function of $s$
for 3D LNPVT.
}
\label{ckiang_volume_exp}
\end{figure}

A  3D    example for $c$ as function of $s$
is reported in
Figure \ref{ckiang_volume_exp} for the LNPVT case
when 
$c_{{\min}}=3.34 $, 
$c_{{\max}}=454.12 $  
and $a$=4.25.

\section{Applications}
\label{applications}
Two applications are presented.

\subsection{An application to chemistry}

\begin{figure}
\begin{center}
\includegraphics[width=10cm]{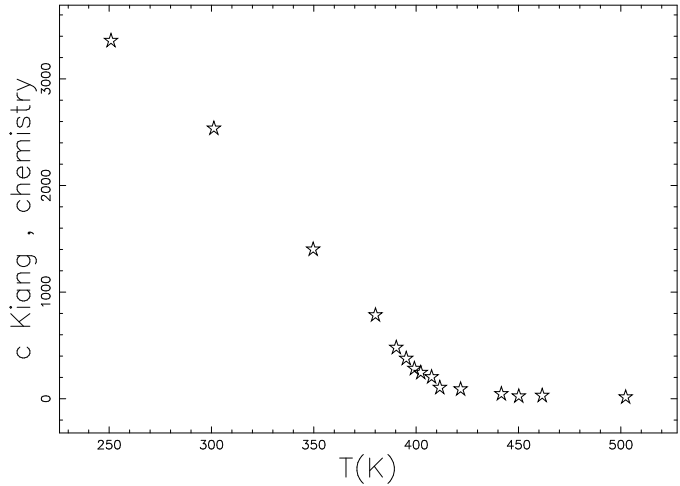}
\end {center}
\caption
{
The Kiang parameter $c$   as function of  T.
}
\label{chemistry_c}
\end{figure}

\begin{figure}
\begin{center}
\includegraphics[width=10cm]{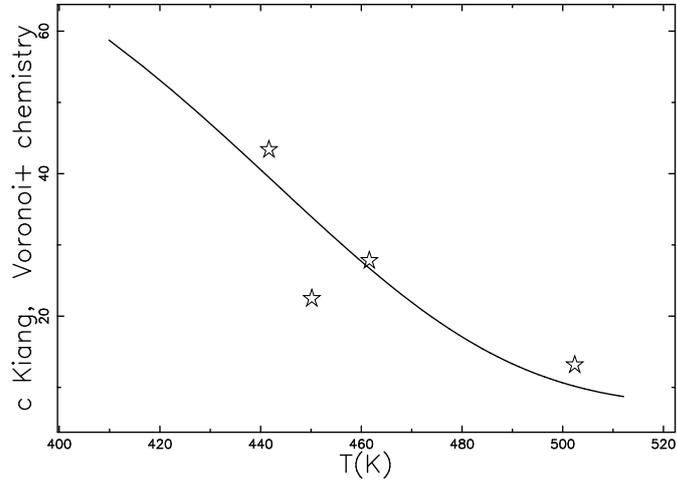}
\end {center}
\caption
{
The Kiang parameter $c$   as function of  T.
}
\label{chemistry_s}
\end{figure}

The local structure of sub- and supercritical ammonia
with $ 250\,K < T < 500\, K$
has been extensively analyzed  by \cite{Idrissi2011}
and Figure \ref{chemistry_c} reports
the $c$ of Kiang  as function of the temperature
for Figure 5 in \cite{Idrissi2011}.
These   values for $c$ of Kiang  are
obtained from Figure 5 in \cite{Idrissi2011}
and
we report as an  example  the deduction of the first couple:
$\sigma_V(\AA^3)$= 4.33 ,$T(K)=250.9$, $\sigma=\frac{4.33}{250.9}
= 0.017$
which means $c=3356$.
Figure \ref{chemistry_s} reports  both
the chemical  and theoretical data:
data with stars as deduced
from Figure 5 in \cite{Idrissi2011}
and full line derived coupling equations (\ref{cs}) and  (\ref{ts}).

To set up the theoretical data
from equation (\ref{cs})
the following  relationship has been used
\begin{equation}
T  = 512\times s ^{0.096} \,K
\quad  .
\label{ts}
\end{equation}

\subsection{An application to cosmic voids}

\begin{figure*}
\begin{center}
\includegraphics[width=10cm]{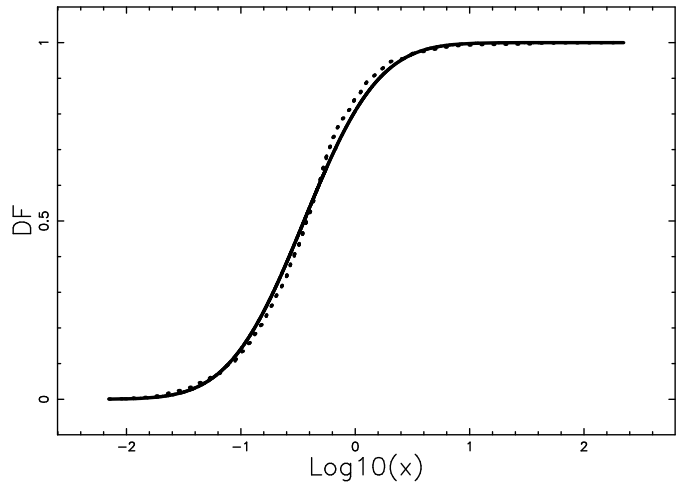}
\end{center}
\caption
{
Empirical DF   of reduced voids  distribution
for  BOSS (dotted line) 
and  lognormal
DF  (full line).
}
\label{lognorm_voids}
\end{figure*}

The  catalog  of the
Baryon Oscillation Spectroscopic Survey (BOSS),
see \cite{Mao2017},
reports  the volume in units of $Mpc^3/h$
of 1228  cosmic voids  where
$h=H_0/100$
and
the Hubble constant, $H_0$,  is   expressed in  $\h0units$.
The  numerical analysis  gives  $c = 0.02$ for the
distribution of the reduced volume of cosmic voids
and
Figure \ref{lognorm_voids} reports
the distribution function (DF)  of the lognormal,
see equation (\ref{dflognormal})
 with parameters as in Table
\ref{lognormalparameters}.

\begin{figure*}
\begin{center}
\includegraphics[width=10cm]{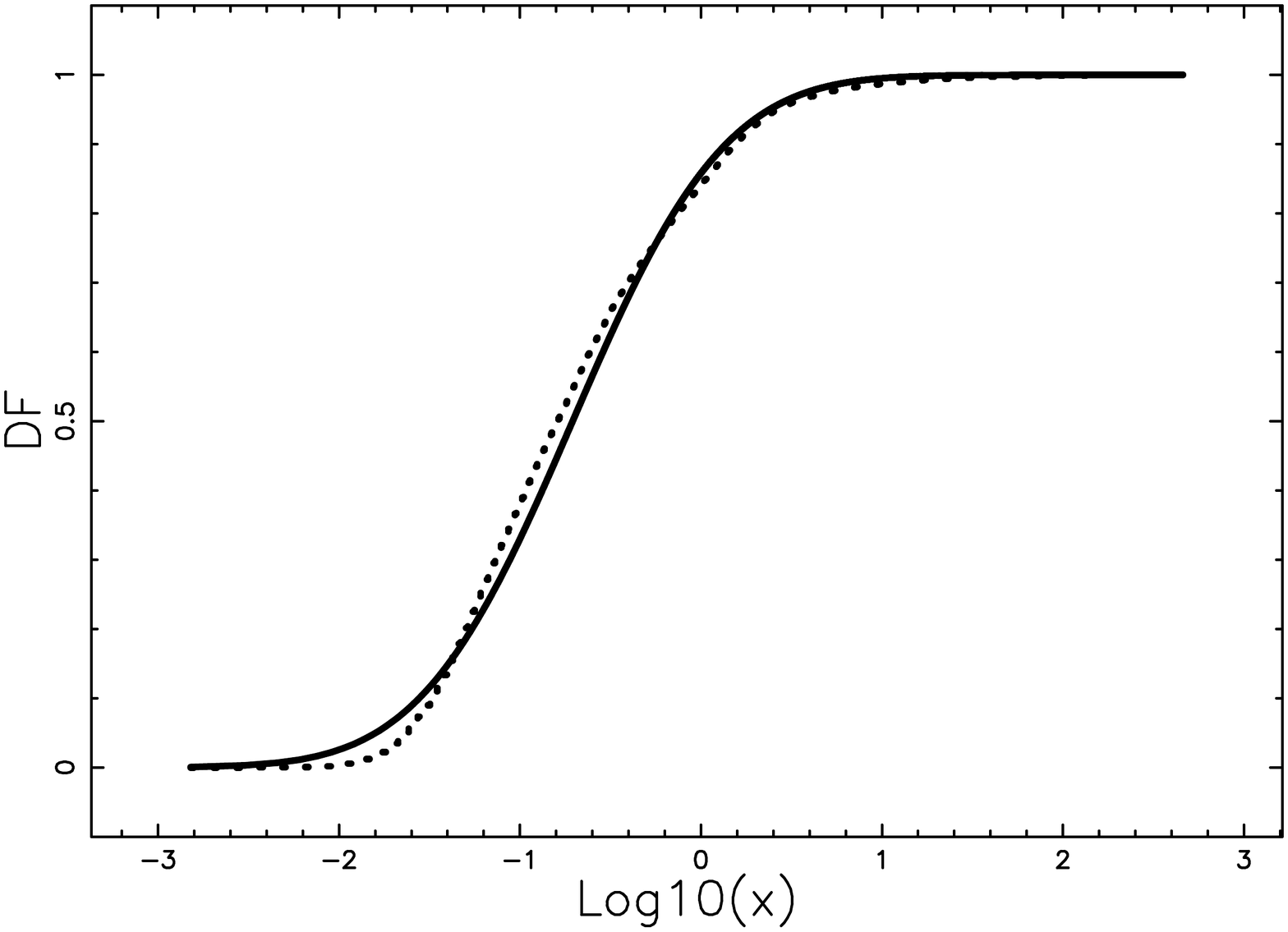}
\end{center}
\caption
{
Empirical DF   of reduced volume  distribution
for 3D LNPVT.
}
\label{theolognorm}
\end{figure*}

These results for the cosmic voids  can  be simulated
analyzing the distribution of the reduced volumes
in 3D tessellation from LNPVT, see Figure \ref{theolognorm},
where we have 
 195112 original seeds, $s=10$  and $p_{hole}= 0.861$.
In this figure, the   lognormal
DF  is the full line and 
the  parameters are as in Table
\ref{lognormalparameters},
where $D$ is  the maximum distance between theoretical and observed $DF$,
and  $P_{KS}$  , is the significance level,   in  the K-S test.

\begin{table}[ht!]
\caption
{
Lognormal parameters for reduced volumes of cosmic voids
and relative LNPVT simulation.
}
\label{lognormalparameters}
\begin{center}
\begin{tabular}{|c|c|c|c|}
\hline
case    & parameters  &    D &   $P_{KS}$                  \\
\hline
astronomical~observations~ & $m= 0.356$; $\sigma$= 1.182  &   0.049    &   0.005 \\
\hline
LNPVT~simulation~ & $m= 0.195 $; $\sigma$=1.523   &     0.055  & 0.003  \\
\hline
\end{tabular}
\end{center}
\end{table}

\section{Conclusions}

We perturbed the 2D seeds  identified by a  Cartesian and
a triangular lattice,  as well the 3D seeds of a Cartesian lattice.
The  probability to have  some holes in the resulting 2D/3D
seeds is introduced.
The area and volume distribution
are modeled by the one parameter Kiang's PDF.
The transition from order to disorder is parametrised by a
geometrical variable,  which regulates the
strength of the perturbation of the ordered seeds.
Two applications are presented:
one to the local structure of sub- and supercritical ammonia
and the second one
to the volumes of cosmic voids.


\end{document}